\documentclass[12pt]{article}
\usepackage[left=2.5cm,top=2cm,right=2.5cm,bottom=2.5cm,nohead]{geometry} 
\usepackage{graphicx}
\usepackage{hyperref}
\usepackage{setspace}
\usepackage[square]{natbib}
\usepackage{amssymb}
\usepackage{verbatim}
\usepackage[usenames,dvipsnames]{color}
\hypersetup{colorlinks=true, citecolor=blue, linkcolor=blue, urlcolor=blue, filecolor=black}

\newcommand{\imloc}{}
\newcommand{\baseloc}{}

\begin{document}

\title{A Practical Implementation of the Bernoulli Factory}
\author{A.C. Thomas\thanks{Department of Statistics, Carnegie Mellon University. Corresponding author; email \href{mailto:act@acthomas.ca}{act@acthomas.ca}} \and Jose H. Blanchet\thanks{Department of Industrial Engineering and Operations Research, Columbia University.}}

\date{\today}
\maketitle

\begin{abstract}
The Bernoulli Factory is an algorithm that takes as input a series of i.i.d. Bernoulli random variables with an unknown but fixed success probability $p$, and outputs a corresponding series of Bernoulli random variables with success probability $f(p)$, where the function $f$ is known and defined on the interval $[0,1]$. While several practical uses of the method have been proposed in Monte Carlo applications, these require an implementation framework that is flexible, general and efficient. We present such a framework for functions that are either strictly linear, concave, or convex on the unit interval using a series of envelope functions defined through a cascade, and show that this method not only greatly reduces the number of input bits needed in practice compared to other currently proposed solutions for more specific problems, and is easy to specify for simple forms, but can easily be coupled to asymptotically efficient methods to allow for theoretically strong results.
\end{abstract}


\doublespacing

\section{Introduction}

First made explicit by \citet{keane1994bf}, a Bernoulli Factory is defined as an algorithm that takes as its input an i.i.d. sequence of Bernoulli random variables with unknown success probability -- call this $p_{in}$ -- and outputs a new sequence of Bernoulli random variables whose success probability, $p_{out}$, is a known function of the input probability. A Bernoulli Factory does not use any approximation for either $p_{in}$ or $p_{out}$, instead obtaining output draws through a stochastic process with two absorbing states, one of which has terminal probability $p_{out}$. The specification of the ``factory function'', by which $p_{out}=f(p_{in})$, is what makes this problem of general interest; the difficulty is in creating a general algorithm that is both implementable and effective with various families of factory functions. 

While previous implementations have been demonstrated to have ``fast'' asymptotic convergence properties under particular conditions \citep{nacu2005fsncfo}, and are special cases of converging super- and sub-Martingale sequences \citep{latuszynski2011seupvrtm}, they are often functionally inefficient at producing output bits for practical problems. Methods based on enveloping Bernstein polynomial approximations, which are naturally approximated by strings of random bits, are optimal at fitting one envelope on a concave or convex function, but extremely inefficient at fitting the other, since successive approximations toward one limit are naturally nested, but not for the other. Our proposed correction to this, a cascading series of envelopes for the ``non-natural'' limit, is an artificial nesting scheme, ensuring that the nesting property is available for both the upper and lower limits to the factory function in a straightforward manner.

We begin by reviewing the roots of the problem, before addressing the Bernstein polynomial approach in Section \ref{s:bernstein}. We then address the implementation of the Cascade Bernoulli Factory method for piecewise-linear concave/convex functions in \ref{s:standard-case}, then demonstrate its potential for more general functions in Section \ref{s:extensions}.

\subsection{Fair Coins and the von Neumann Algorithm}

The prototypical problem comes from \citet{vonneumann1951vtucrd}, which seeks to generate a ``fair coin'', or a draw from a $Be(0.5)$ random variable, from an i.i.d. sequence of Bernoullis with unknown success probability $p$ -- that is, $f(p)=0.5$ for all $0<p<1$. The corresponding stochastic process has three states, labelled ``yes'', ``no'' or ``continue'':

\begin{itemize}

\item Begin in state ``continue''.

\item Take two draws from the input sequence. The possible outcomes are grouped as $\{00, 01, 10, 11\}$.

\item While the outcome is 00 or 11, remain in state ``continue''. Discard these bits and replace them with two new draws.

\item If the outcome is 10, output ``yes''; if the outcome is 01, output ``no''.

\end{itemize}

Because both 01 and 10 have probability $p(1-p)$ or occurring, there is equal probability of the outcome being a ``yes'' or a ``no'', and as a result, the outcome can be likened to the flip of a fair coin. The running time of this method is two times a Geometric random variable with success probability $2p(1-p)$, so that the expected number of input bits required is $\frac{1}{p(1-p)}$.

A similar process can be conducted to turn a series of fair coins into a coin with any success probability $p_{out}$, by noting that a uniform random variable can be produced through the representation

\[ U = \sum_{i=1}^{\infty} 2^{-i} X_i \]

\noindent where $X_i \sim Be(0.5)$. However, a finite number of bits will be needed if the outcome of interest is specified as

\[ Y = \mathbb{I}(U < p_{out}). \]

The stochastic process has an unbounded number of states, but is as simple to specify as the standard von Neumann example:

\begin{itemize}

\item Begin with $n=1$. 

\item At each stage $n$, set $U_n = \sum_{i=1}^{n} 2^{-i} X_i$. Note that $U_n \leq U_{n+k}$ for all $n$ and $k$.

\item If $U_n > p$, then $U > p$ as well; output ``no''.

\item If $U_n < p - 2^{-n}$, then no matter what the remaining inputs are, $U < p$; output ``yes''. 

\item Otherwise, add one more digit to the expansion and repeat the previous three steps.

\end{itemize}


These algorithms each converge in geometric time, with rates proportional to the target probability $p_{out}$. While neither requires a sophisticated implementation in order to produce a correct output, these methods suggest a general trend: that methods that produce absorbing states of simple Markov chains are powerful methods of transforming random bits without a loss of information, merely efficiency.

\subsection{From Simple Alchemy to the Full Factory}

The problem explored by \citet{keane1994bf} works on the principle that the input and output success probabilities are possibly unknown, but a function that defines their connection is fully specified. The case where

\[ f(p) = \min\left(cp, 1-\epsilon\right),\ c>1,\ \epsilon<1,  \] 

\noindent henceforth referred to as the ``elbow function'', is of particular interest to applications in exact sampling of Markov chains \citep{asmussen1992sditp, hobert2004mrpamcmcaps, blanchet2005esramc, hobert2006umcctaipd, blanchet2007esaesvrabf, flegal2012exact}, as this function represents a ratio in general rejection sampling schemes for draws from the stationary distributions of Markov Chains. (In the case where $c<1$, the problem is trivial: the representation $X \sim Be(c)*Be(p)$ immediately produces the desired result.)

To make practical use of this, the solution proposed by \citet{nacu2005fsncfo} uses a pair of Bernstein polynomial forms to approximate $f(p)$ from above and below. The standard Bernstein polynomial approximation to a function is defined as 

\[ f_n(p) = \sum_{k=0}^n {n \choose k} f\left(\frac{k}{n}\right) p^k \left(1-p\right)^{n-k}; \]

\noindent their usefulness comes about in this problem because the probability of any one of the sequences of $k$ ones and $n-k$ zeros from $n$ $Be(p)$ random variables is $p^k (1-p)^{n-k}$. If these approximations both converge to the target function in the limit, then a draw from the target distribution $Be(f(p))$ can be obtained in finite time. 

The theoretical properties of this method have been described in detail by previous authors, though the practical implementations of these solutions are concerned more with theoretical tractability than flexibility or applicability to real problems. \citet{keane1994bf} showed that a factory function must be continuous and satisfy the inequality

\[ \min (f(p), 1-f(p)) \geq \min (p, 1-p)^n \]
for some $n \geq 1$ and all $0<p<1$; \citet{nacu2005fsncfo} established conditions for ``fast'' simulation, such that the required number of input bits decays exponentially, and proved that this held for all real analytic functions bounded away from 0 and 1. 


\section{Bernstein Polynomial Expansions and Set Approximations}\label{s:bernstein}

Bernstein polynomials are a set of basis functions defined on the interval $[0,1]$. A Bernstein approximation of order $n$ contains a total of $n+1$ functions of the form $p^k(1-p)^{n-k}$, which is also the individual probability of any single sequence containing $k$ ones and $n-k$ zeros. Indeed, it is trivial to show that the Bernstein polynomial approximation is the expected value of the function under a Binomial random variable:

\[ K \sim Bin(n,p);\quad Ef\left(\frac{K}{n}\right) = \sum_{k=0}^n {n \choose k} f\left(\frac{k}{n}\right) p^k \left(1-p\right)^{n-k} = f_n(p). \]

Using this result, a concave function will always be greater than any of its finite Bernstein polynomial expansions; by Jensen's inequality, $f(p) = f(EK/n) \geq Ef(K/n) = f_n(p)$. Conversely, a convex function will always be less than any of its Bernstein expansions.

The general method proposed by \citet{nacu2005fsncfo} works with the use of two approximating functions, one from above, one from below, and ties these directly to the probabilities of observing particular bit strings. With a slight change in notation, consider the following formulation:

\begin{itemize}

\item Define a series of functions $a^{n}(p)$ that approximate the target function $f(p)$ from below (that is, for all $n$, $a^n(p) \leq f(p)$ and $\lim_{n \rightarrow \infty} a^n(p) = f(p)$). Define also another series of functions $b^n(p)$ to approximate $1-f(p)$, also from below; by construction, $a^n(p)+b^n(p) \leq 1$. These functions are used when the total length of the input bit string is $n$.

\item Let $A_n$ be a set of bit-words of length $n$, and $A_{n,k}$ be the subset of $A_n$ with exactly $k$ ones; likewise with $B_n$.

\item The Bernstein polynomial approximations $a_n^n(p)$ and $b_n^n(p)$ yield natural derivations for subsets $A_{n,k}$ and $B_{n,k}$. Since

\[ a_n^n(p) = \sum_{k=0}^n {n \choose k} a^n\left(\frac{k}{n}\right) p^k \left(1-p\right)^{n-k}, \] 
we introduce 

\[ A_n^n(p) = \sum_{k=0}^n \left\lfloor {n \choose k} a^n\left(\frac{k}{n}\right) \right\rfloor p^k (1-p)^{n-k}; \]
and define a set $A_{n,k}$ containing $\lfloor {n \choose k} a^n(\frac{k}{n}) \rfloor$ distinct $n-$length bit strings with $k$ ones. From those bit strings that remain, choose $\lfloor {n \choose k} b^n(\frac{k}{n}) \rfloor $ to form $B_{n,k}$ (noting that the probability of observing any one bit-string is $p^k(1-p)^{(n-k)}$.)

\item Ensure that $A_n^n(p) \leq f(p)$ and $B_n^n(p) \leq 1-f(p)$. For whatever sequence of functions is used, also ensure that $A_{n+m,k} \geq \sum_{j=0}^m {m \choose j} A_{n,k-j}$, so that any lower-length bit string in $A_n$ may also be in $A_{n+m}$.

\item Collecting the remaining unaccounted items, define $C_{n,k}$ to be all those $n-$length bit strings with $k$ ones that were not included in $A_{n,k}$ or $B_{n,k}$.

\end{itemize}

Using these tools we can now build the Bernoulli factory for a great number of classes of functions; details of the convergence properties of this method for various proposed envelope functions are addressed by \citet{keane1994bf} and \citet{nacu2005fsncfo}.


\subsection{Example: $f(p)$ is Constant or Linear}

With a linear factory function $f(p)=c + hp$, it is clear that the standard Bernstein polynomial expansion is identical:

\begin{eqnarray}
f_n(p) & = & \sum_{k=0}^n {n \choose k}\ (c + h\frac{k}{n}) \ p^k (1-p)^{n-k} \\
& = & c + \frac{h}{n} \sum_{k=0}^n {n \choose k}\ k \ p^k (1-p)^{n-k} \\
& = & c + \frac{h}{n} EK = c + \frac{h}{n}np = c+hp. 
\end{eqnarray}

As a result, the function can be used as both an upper and a lower envelope, so long as $0<f(p)<1$ for all $0<p<1$. This means that for those cases where ${n \choose k}\ (c + h\frac{k}{n})$ is an integer, $C_{n,k}$ is empty, and the algorithm will terminate if $k$ ones are observed. For the case when it is not an integer, there will be only one member of $C_{n,k}$. As a result, the survival function is bounded above by the simple expression

\[ P(T>n) \leq \sum_{i=0}^n p^i (1-p)^{(n-i)}. \] 

\begin{table}
\begin{center}
\begin{tabular}{cccc}
\hline
\hline
k,n=2 & $A_{2,k}$ & $B_{4,k}$ & $C_{2,k}$ \\
\hline
0 & & & 00 \\
1 & 10 & 01 & \\
2 & & & 11 \\
\hline
\hline
k,n=4 & $A_{4,k}$ & $B_{4,k}$ & $C_{4,k}$ \\
\hline
0 & & & 0000 \\
1 & 0010,1000 & 0001,0100 & \\
2 & {\bf \textcolor{red}{0011}}, 1010, 1001   &    0101, 0110, {\bf \textcolor{red}{1100}} & \\
3 & 1110, 1011 & 1101, 0111 & \\
4 & & & 1111 \\
\hline
\hline
\end{tabular}
\end{center}
\caption{The bit strings needed for a Bernoulli factory with function $f(p)=0.5$ in the first two steps. This algorithm is slightly more efficient than von Neumann's algorithm for manufacturing fair coins. \label{table:von-neumann}}
\end{table}

To demonstrate, consider the von Neumann problem again, so that $f(p) = 0.5$ for whatever $p$. Setting $a^n(x)=b^n(x)=0.5$ for all $x$, for any $n$ it is clear that the size of any $C_{n,k}$ is either 0 or 1, if ${n \choose k}$ is even or odd respectively. In particular, Table \ref{table:von-neumann} describes the cases where $n=2$ and $n=4$, and a potential distribution of bit strings over sets. The sequences in the respective $A$, $B$ and $C$ sets once again represent ``output 1'', ``output 0'' or ``add more bits'', but under the Bernstein construction, there are two additional bit strings that will terminate the algorithm where von Neumann would not: 0011 and 1100. Note that the ``descendants'' of $A_{2}$ also appear in $A_{4}$ and indeed all $A_{n}$ beyond $n=2$.

In practice, it is not necessary to construct this table for all $n$, or even to select a particular partitioning of all bit-strings, only to note the size of the sets themselves. As we demonstrate, this is done by ensuring that for any string in $A_n$, all of its ``descendants'' obtained by adding any $m$-length bit string are members of $A_{n+m}$, and similarly for $B_n$ and $B_{n+m}$.


\section{The Canonical Case: $f(p)$ is Piecewise Linear, and Concave (or Convex)}\label{s:standard-case}

The key motivating problem for the practical use of the Bernoulli factory is the aforementioned elbow function made famous by \cite{keane1994bf} and proposed in connection with \citet{asmussen1992sditp},

\[ f(p) = \min(cp, 1-\epsilon),\ c>1,\ \epsilon<1, \] 

\noindent which is concave. Due to Jensen's inequality, the Bernstein polynomial approximation to this function will always be less than the target function, so the lower bound function is simply $a^n(p)=f(p)$.

The upper envelope function is considerably more difficult to design effectively. A single function cannot be used, as the Bernstein approximation to a concave function will increase as the length of the bit string increases. The envelope functions chosen by \citet{nacu2005fsncfo} are functions of the bit-string $n$ and are sufficient to prove the convergence properties of the algorithm under particular constraints, but are markedly inefficient at producing output draws; \citet{flegal2012exact} show that a minimum of $2^{16}$ bits are required for the function $f(p)=\min(2p, 0.8)$. The capability of current computing hardware to efficiently calculate Bernstein expansions for functions above this count renders the algorithm not only inefficient in terms of input, but hopeless for practical implementation.

The alternative specification of \citet{flegal2012exact} changes the problem slightly by specifying a new objective function that is twice-differentiable, but still linear on the domain $[0,(1-\epsilon)/c]$, as \citet{nacu2005fsncfo} proved that certain twice-differentiable factory functions converge at a rate of $1/n$ (which yields an infinite expected time to termination, even though the termination time is finite.) The number of bits needed is considerably reduced from the original case, but still requires a minimum of many hundreds of bits to operate on these target functions. 

This expansion does have nice theoretical convergence properties across the entire domain, but practical applications of this function typically take place when $p$ is small (as demonstrated by \citet{flegal2012exact}); the flattening-out of the function is to ensure continuity rather than as a particular range of interest. Because convergence at small $p$ is typically the goal, we choose our sampler to visibly converge well in this range.

Rather than create a new functional form to add to the target function in creating an upper envelope, albeit one that would not noticeably affect the output of the algorithm for known inputs, our approach is to use the existing function to create a series of cascading envelopes that will converge to the target function from above (approaching $1-f(p)$ from below). In particular, the manner in which the functions cascade is governed by their own Bernstein polynomial expansions, and the sequence can easily be constructed to converge to the target function $f(p)$ in the limit. (The rate at which this convergence occurs will depend on the chosen envelopes; there must, however, exist an asymptotically ``fast'' version by coupling this to another series of envelopes that must exist; see Section \ref{s:theory}.)

We require a series of ``checkpoints'' $\{m_1, m_2, ...\}$ at which the envelopes will be constructed and used. As \citet{nacu2005fsncfo} point out, it is trivial define a partition $(A_{n+\Delta n}, B_{n+\Delta n}, C_{n+\Delta n})$ by starting with a partition $(A_{n}, B_{n}, C_{n})$ and adding all possible $2^{\Delta n}$ bit strings to each member of each set; this freedom allows us to minimize computation by choosing a smaller set of tests to conduct. The choice of checkpoints can be defined in any number of ways but may also be chosen to minimize the running time of the algorithm.

\begin{figure}
\begin{center}
\includegraphics[width=0.4\linewidth]{\imloc 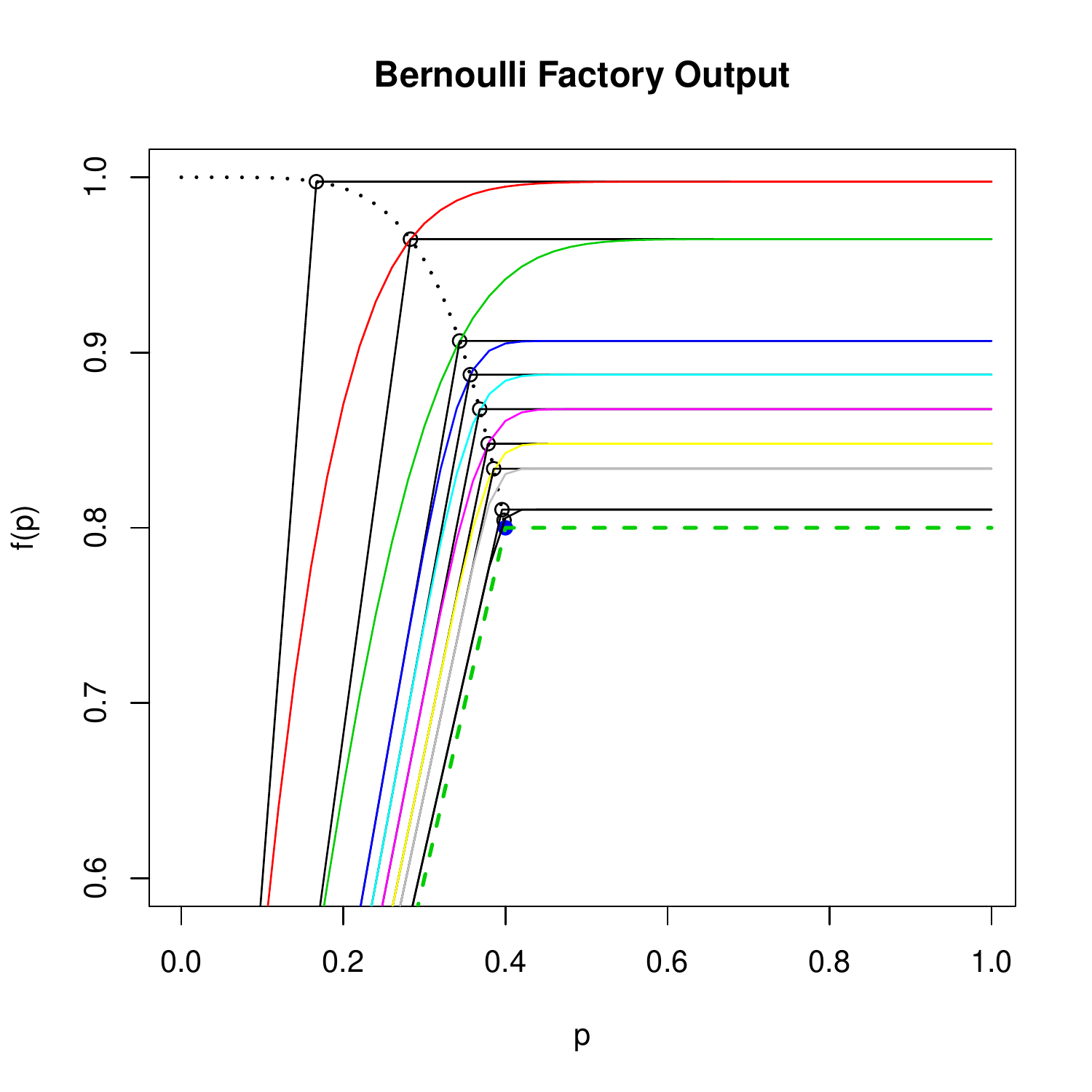}\includegraphics[width=0.4\linewidth]{\imloc 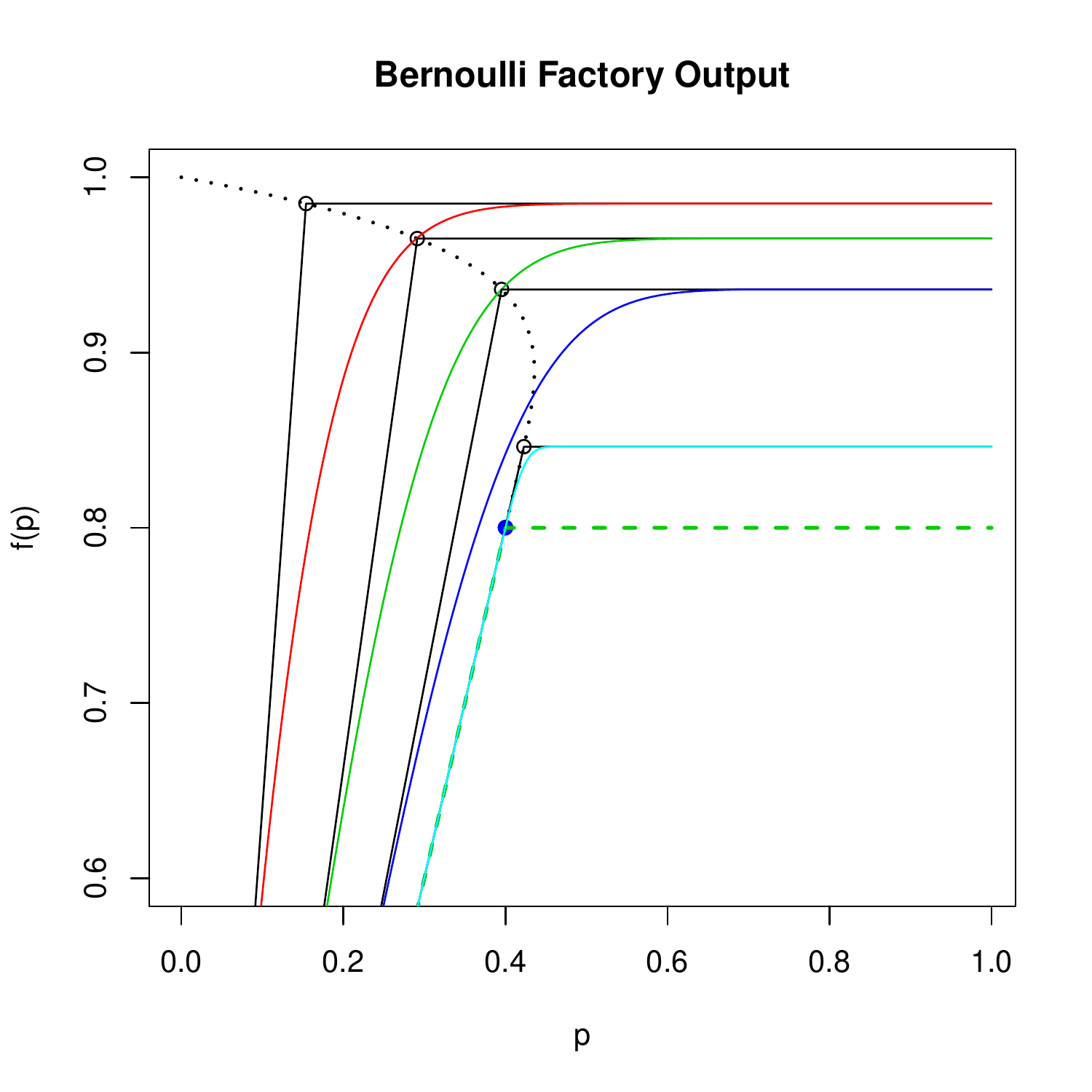}
\caption{Two methods for generating upper envelope functions for $f(p)=\min(2p, 0.8)$; each successive function is defined by the intersection of a curve with the previous Bernstein approximation. Left, the elbow points are generated with a simple polynomial descent (dotted line), as each successive curve approaches the target function (dashed line). Right, the descent curve is a fifth-degree polynomial in $y$ with respect to $x$, with derivative at the main elbow point matching that of the target function, designed to minimize the distance between the Bernstein polynomial and the target function for smaller values of $p$.\label{cascade}}
\end{center}
\end{figure}

The method takes the following steps:

\begin{itemize}

\item[a)] Choose a potential series of functions that converge toward $f(p)$. As shown in Figure \ref{cascade}, we select a series of elbow functions whose elbow points lie along a preset curve (two such curves are demonstrated). 

While any curve can be chosen that will produce a nested set of envelopes, the second curve, a fifth-degree polynomial whose slope at the elbow point matches the elbow function, fits the diagonal ascent of the function more tightly and will therefore be more efficient when the input bits have very small probability. For this function, higher-degree polynomials will flatten out the curve with respect to the elbow and force each Bernstein approximation to be closer to these small values.

\item[b)] Choose an initial elbow point along this curve, and initial bit-string length $m_1$. In this case it is simple to verify that $1-b_{m_1}^{m_1}(p) > f(p)$ for all $p$ by evaluating $b_{m_1}^{m_1}$ at the target function's elbow point $(1-\epsilon)/c$, as the Bernstein expansion's concavity ensures that we only need check the connecting points of the piecewise linear $f(p)$.

\item[c)] Retrieve $m_1$ bits from the input bit stream and set $k_1$ to equal the number of ones. Note the sizes of each subset $A_{m_1,k_1}$, $B_{m_1,k_1}$ and $C_{m_1,k_1}$. If desired, one can generate the actual corresponding bit strings, but this is unnecessary to run the algorithm itself.

\item[d)] If the bit string memberships of each groups have been specified exactly, note which of the subsets contains the observed string; if not, generate a trinomial random variable with probabilities proportional to $(|A_{m_1,k_1}|, |B_{m_1,k_1}|, |C_{m_1,k_1}|)$. Terminate the algorithm with output 1 or 0 if this trinomial is in each of the first two bins; If not, continue.

\end{itemize}

For all subsequent steps indexed by $i$: 

\begin{itemize} 

\item[e)] Choose the next elbow point to be at (or below) the point where the previous Bernstein approximation $1-b_{m_{i-1}}^{m_{i-1}}(p)$ intersects the elbow point curve; this ensures that the next envelope is less than or equal to the previous envelope's Bernstein approximation. Additionally, choose a value $m_{i} > m_{i-1}$ such that $1-b_{m_i}^{m_i}(p) > f(p)$, and so that sufficient room is left for future iterations (since if the envelope is too close to the target, an extreme value of $m$ will be needed to produce a Bernstein expansion greater than the target function.) This is the next upper envelope in the cascade; by construction, it is less than the previous envelope and produces a Bernstein expansion that is less than its precursor. Figure \ref{cascade} contains two examples, one designed to approach the function relatively uniformly (a quadratic function) and one designed to hew close to the ascending line (a rotated fifth-degree polynomial.) 

This particular method of descent is chosen for this function because it guarantees this nesting behavior while being easy to understand and implement; aside from this condition, there are no particular restrictions on the method by which envelopes are produced. A default descent function merely makes it more automated to the end user, though this can change depending on the desired convergence properties of the method. Figure \ref{cascade} shows that a function that quickly approaches the ascending portion of the function will yield more early terminations, by closing the distance between functions; the downside is that if the algorithm does not terminate quickly,

\item[f)] Retrieve $m_i - m_{i-1}$ bits and add them to the current bit string; let the number of ones equal $k_i$. Calculate the sizes of sets $A_{m_i,k_i}$, $B_{m_i,k_i}$ and $C_{m_i,k_i}$ for $k_i$. For each element in $A_{m_{i-1},k_{i-1}}$, there are ${m_i - m_{i-1} \choose k_i - k_{i-1}}$ elements that are produced by adding $(m_i - m_{i-1})$-bit strings with $(k_i - k_{i-1})$ ones; since these would have produced termination in the prior step, remove these from $A_{m_i,k_i}$; similarly, remove the redundant descendants from $B_{m_i,k_i}$.   

\item[g)] Generate a trinomial random variable with probabilities proportional to the set sizes,

\[ Pr(1,0,\mathrm{continue}) \propto (|A_{m_i,k_i}|, |B_{m_1,k_i}|, |C_{m_1,k_i}|). \] 
Terminate the algorithm with output 1 or 0 if this trinomial is in each of the first two bins, and repeat these steps if not for the next checkpoint $m_{i+1}$.

\end{itemize}


In general, these set sizes can all be pre-calculated for as many cascade steps $i$ as is desired, and for all possible $k$; the enumeration of bit strings into sets is not required, since if the algorithm continues at each $i>1$, it must be that the previous bit string belonged to group $C$.

\subsection{Practical Justification and Improvement}\label{s:practical}

For any envelope cascade, series of checkpoints, and an input probability value, one can calculate the marginal probability of each output bit's manifestation simply by adding up the total probability of all bit strings in each set $(A_m, B_m, C_m)$; this is done on the graphs for demonstration purposes, but can also be produced in table form in order to gauge the running time of any particular envelope set.

\begin{table}
\begin{center}
\begin{tabular}{ccccccc}
\hline
\hline
$p_{in}=0.01$ & $m_i$ & elbow x & elbow y & P(output 1) & P(output 0) & P(continue) \\
\hline
1 & 20 & 0.1539 & 0.985 & $0.02 - 1.537\times 10^{-14}$ & 0.9309 & 0.04898 \\
2 & 21 & 0.2912 & 0.965 & $0.02 - 1.536\times 10^{-14}$ & 0.9612 & 0.01881 \\
3 & 222 & 0.3953 & 0.936 & $0.02 - 8.573\times 10^{-15}$ & 0.9712 & 0.00882 \\
4 & 1223 & 0.4228 & 0.8463 & $0.02 - 1.422\times 10^{-16}$ & 0.9799 & 1.575$\times 10^{-5}$ \\
\hline
\hline
\end{tabular}
\end{center}
\caption{Convergence characteristics for the execution of a Cascade Bernoulli Factory with function $f(p)=min(2p, 0.8)$. Because this function is concave, the lower curve approximation is far more efficient than the upper curve.\label{tab:convergence}}
\end{table}

Consider the envelope cascade from Figure \ref{cascade}, which has a target elbow function with $(c,\epsilon)=(2, 0.2)$, using the right-hand descent function. For a given $p_{in}$, the probabilities of belonging to each group at each stage are calculated and listed in Table \ref{tab:vs-flegal}. Note that these probabilities represent the ultimate termination probability of the algorithm, not the conditional probability of the outcome given that said number of steps was necessary. In this case, survival past the first round all but secures an eventual output of zero; the price paid for perfection of the algorithm is the high number of additional bits needed to guarantee that result.

A lower bound on the expected value for the number of bits required per output can be estimated with the well-known formula $EN = \sum_{n=0}^{m_{max}} P(N>n)$, since the marginal probability of non-termination is equal to the size of $C_m$. Note that in this example, we have engineered a four-tiered envelope structure, and in the last tier shown we have greatly increased the number of input bits, as well as chosen a closer point to the target on the descent curve to increase the probability of termination within these steps; still, if the factory were to be run one million times, we can expect that the algorithm would not terminate at this point in 16 of them. If this is the case, the number of required bits at this point would skyrocket, as we have pinned ourselves into a situation where a vast increase in bits would be necessary to keep the algorithm valid; we call this ``termination by gluttony'' to distinguish from a standard exhaustion of bits due only to the main cascade. 

This is as much a feature of design as any other. Optimizing this factory for practical use takes several directions:

\begin{itemize}

\item Does the user wish to optimize the method with respect to a single input probability, or across a range? The above table can be used for various input probabilities, though how to combine them appropriately will depend on the problem being considered. (See Section \ref{s:extensions} for an example.)

\item Can we use the input bits twice? For a particular factory function, if the bit string input probability is fixed but unknown, we can estimate $\widehat{p} = \frac{1}{N}\sum_{i=1}^N Y_i$ and choose an envelope cascade that quickly approaches this point in the function, since only the output speed is associated with the cascade, not the output probability.

\item Given that the algorithm is not guaranteed to terminate at the final terrace, do we wish to minimize the probability of not terminating before the massive expense, at the cost of an increased number of bits in those cases that do terminate? This may prove to be prudent given that our algorithm is not interruptable \citep{fill1998iafpsvmc} -- that is, we can not terminate any run without introducing bias into the resulting bit stream, since as the iterations progress, we are far more likely to obtain a zero than a one.

\item This should also cause a user to consider whether the number of available input bits is comparatively small, in which case the probability that the algorithm would terminate due to lack of supply would far exceed the probability of termination by gluttony. 

\end{itemize}

Practical improvement over \citet{nacu2005fsncfo} is clear, for an algorithm whose design was theoretically elegant but known to be impractical. The method also yields considerable practical improvement over that proposed by \citet{flegal2012exact} in terms of both the minimum number and the expected number of input bits required for a single output, mainly because our method for this function always has $|B_{n,0}| = 1$ (as $b^n(0)=1$), if the first considered bit sequence contains no ones, the algorithm will output a zero on the first round. This result is perfectly valid if the envelope is shown to be strictly greater than the target function, which requires that the number of input bits exceed some minimum value.

\begin{table}
\begin{center}
\begin{tabular}{cccc|ccc}
\hline
\hline
& Cascade & Method & & & Best & Alternative \\
c & min(bits) & E(bits) & sd(bits) & min(bits) & E(bits) & sd(bits) \\
\hline
2 & 20 & 66 & 512 & 256 & 562.9 & 2104.6 \\
5 & 100 & 246 & 1215 & 2048 & 2439.8 & 7287.6 \\
10 & 200 & 614 & 1851 & 8192 & 10373 & 54836 \\
20 & 400 & 1410 & 3047 & 32768 & 43771 & 390800 \\
\hline
\hline

\end{tabular}
\end{center}
\caption{Properties of the Cascade Bernoulli Factory against that proposed by \citet{flegal2012exact}.\label{tab:vs-flegal}}
\end{table}

For $\epsilon=0.2$, the number of input bits for a standard implementation of the Cascade Bernoulli Factory is given in Table \ref{tab:vs-flegal}, comparing this method against the \citet{flegal2012exact} implementation for various multipliers $c$, over $10^4$ trials, with $p_{in}=0.01$. The termination probability was chosen to be 1 in $10^6$ for most examples, though for the case when $c=2$, it is easy to find a ``small'' value of $n$, on the order of 10000, where the probability of continuing past the last pre-calculated envelope is so small that the computer cannot distinguish it from zero. 

The minimum number of draws shown in this table is not a strict property of the method, but simply a consideration to be made in the choice of envelopes, since there needs to be a comfortable distance between each Bernstein expansion and the target function so that future steps do not require a vast number of additional input bits. The fewer bits that are required at the first checkpoint, the less this distance will be, and the more bits will be required in further steps. Likewise, choosing a higher number of bits and a closer envelope function will decrease the probability that a comparably large number of bits will be required for the algorithm to terminate, but greatly increase the number of bits required if termination does not occur quickly.


\subsection{Theoretical Justification}\label{s:theory}


It is trivial to show that there exists a sequence of envelopes that converges to the factory function as the number of bits increases. Consider the case of a concave factory function, so that the ``difficult'' envelopes come from above.

\begin{itemize}

\item For any fixed envelope $b^n(p)$ and Bernstein polynomial $b^n_n(p)$, for any $\delta > 0$ there exists some $n$ such that $b^n(p) - b^n_n(p) < \delta$ for all $0<p<1$, by the convergence properties of Bernstein polynomials. In the concave case, it is immediate that all $m>n$ also satisfy the condition.

\item There exists a point on the descent function that is either equidistant from any chosen envelope and the factory function, or closer to the function than the envelope. Choose a new envelope $k(p)$ that intersects this point, and set the previous $n$ such that $f(p) < k(p) < b^n_n(p) < b^n(p)$ for all $0<p<1$. Choose a bit count $n_2$ such that $f(p) < k_{n_2}(p) < k(p)$ for all $0<p<1$. 

\item With every successive iteration, the envelope sequence covers half the distance to the target factory function. There must therefore exist a sequence of functions such that $\lim_{k \rightarrow \infty} h^{n_k}_{n_k}(p) - f(p) = 0$.

\end{itemize}

The existence of an asymptotically ``fast'' algorithm is similarly easy to prove: 

\begin{itemize}

\item By \citet{nacu2005fsncfo}, there exists a series of lower and upper envelopes $(g^n(p), h^n(p))$ for which $Pr(N>n) \leq Ce^{-cn}$ for some $(C,c)$. \footnote{The upper envelope functions proposed by \citet{nacu2005fsncfo} for the elbow function take the form $h^n(p) = \min(cp, 1-\varepsilon) + C_1\sqrt{\frac{2}{n}}\max \left(p-\left(\frac{1}{2}-3\varepsilon\right), 0\right) + C_2 \exp\left(-2\varepsilon^2n\right)\max\left(p-\frac{1}{9}, 0\right)$ and are constructed to converge to the target function from above at exponential rates for small $p$ and polynomial rates for other $p$; $C_1$ and $C_2$ are constants required to be large enough so that the function produces valid envelopes. This also requires a very large number of bits for the first acceptable envelope -- $2^{16}$ is often quoted for $c=2$ and $\varepsilon=0.2$, which is still overly burdensome for current hardware.} 

\item For any Cascade envelope $b^n(p)$, define $b^{m}(p) = b^n(p)$ for all $m>n$. If there exists an $m$ such that $h^m(p)$ that is less than $b^m(p)$ for all $0<p<1$, then there must be some $m'>m$ for all $0<p<1$ such that 

\[ f(p) < h^{m'}(p) < b^{m'}_{m'}(p) < b^{m'}(p) \]
as $h^{m'}(p) \rightarrow f(p)$ and $b^{m'}_{m'}(p) \rightarrow b^{m'}(p)$ as $m' \rightarrow \infty$.

\end{itemize}

It is trivial to show that these conditions are satisfied for the Nacu-Peres upper envelope and the Cascade, since the latter has greater slope than the former at $p=0$ where the curves meet. The only catch is that the number of input bits required will be extremely large. For the example in the above table, four envelopes were produced with cascading envelopes. To produce a fifth under the Nacu-Peres criteria, a startlingly high value of $n_5=2^{37}$ is required for the envelope $h_{n_5}(p)$ to first fits between $f(p)$ and the fourth upper envelope $b_{1223}^{1223}(p)$; from this point on, additional envelopes $h_{n_i}$ can be produced that fit below the previous envelope and the target function by doubling each successive $n$. With over 100 \textit{billion} bits required, this would not be at all feasible in practice, but as we have already passed the point of feasibility of executing the original factory operation on existing hardware, this is immaterial. We seek only to establish that there is a choice of envelopes for which the method has a ``fast'' convergence rate.

\section{Extending to General Convex or Concave Functions}\label{s:extensions}

As defined, the cascading envelope method works identically for general convex and concave functions beyond the simple piecewise linear construction we have used so far. The only difference is that the method for guaranteeing that the Bernstein expansion for the upper envelope is greater than the target function is not as simple as checking a finite number of points. A numerical method may need to be used to guarantee that the envelope does not cross the target function.

\subsection{The Smoothed Elbow, Revisited}\label{s:fh-redo}

The proposed factory function of \citet{flegal2012exact} was chosen partly for its asymptotic properties. The factory function is

\[ f(p|c,\epsilon, \delta) = \mathbb{I}\left(p < \frac{1-\epsilon}{c}\right) cp + \mathbb{I}\left(p \geq \frac{1-\epsilon}{c}\right)\left( (1-\epsilon) + q\left( p - \frac{1-\epsilon}{c} \right) \right) \]
where 

\[ q(p|c,\epsilon, \delta) = \delta \int_0^{cp/\delta} e^{-t^2} dt \]
which is twice differentiable, with $|f''(p)| \leq C = c^2\frac{\sqrt 2}{\delta \sqrt e}$. The authors choose $a^n(p) = f(p)$ as before, and 

\[ b^n(p) = f(p) + \frac{C}{2n} \]
for the upper envelope. This has the property of converging at a rate of $\frac{1}{n}$ across the whole function.

We can improve upon this in two ways. First, we can choose an envelope function that equals zero at zero, so that small values of $p$ will have quick rates \citep{nacu2005fsncfo}, while still respecting the condition on the second derivative; the function

\[ b^n(p) = f(p)*(1 + \frac{C}{2n(1-\epsilon)}) \]
maintains this condition. Second, once this is in place, we can preface this sequence with looser envelopes at lower values of $n$ that will ensure earlier termination without interfering with the original sequence.

\begin{figure}
\begin{center}
\includegraphics[width=0.5\linewidth]{\imloc 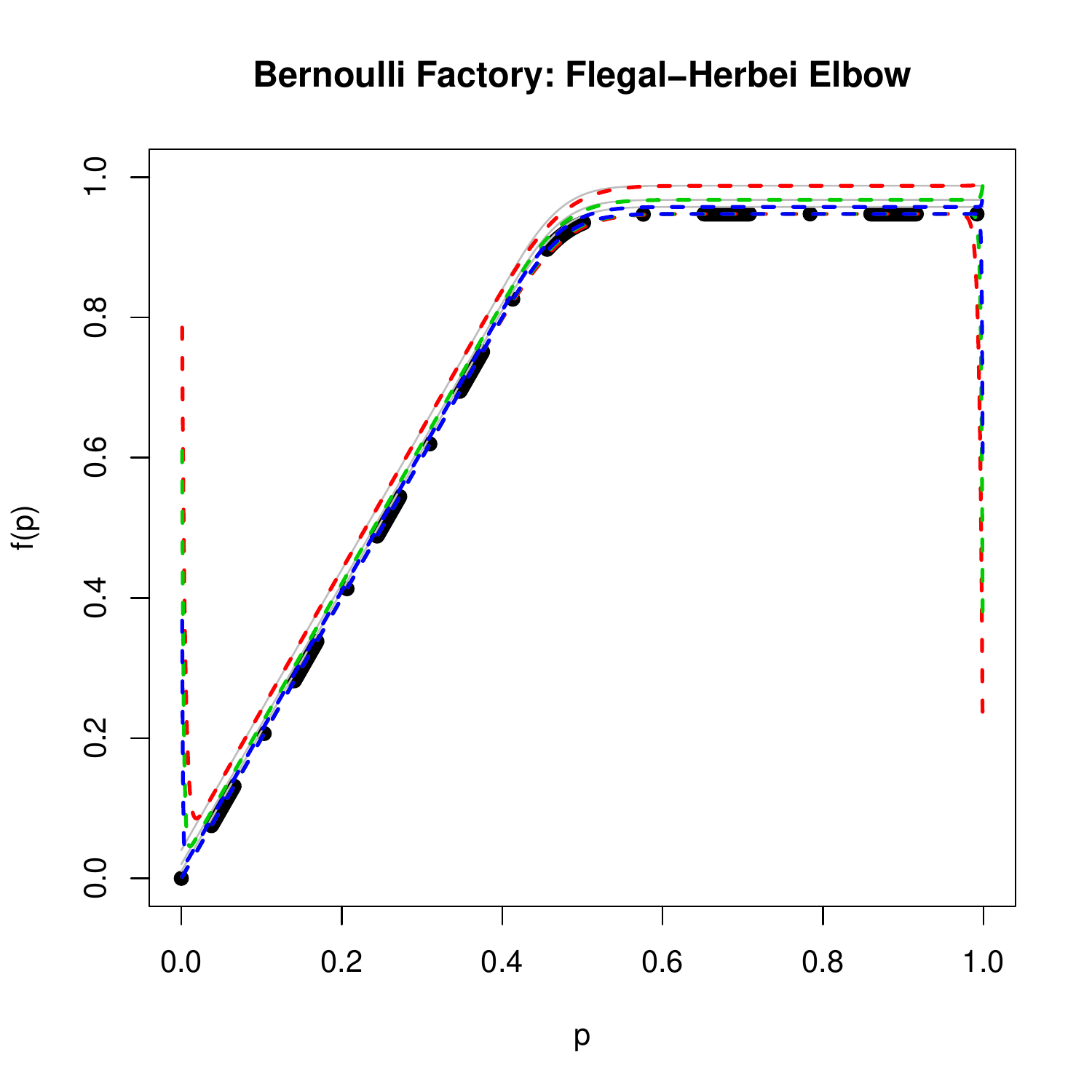}\includegraphics[width=0.5\linewidth]{\imloc 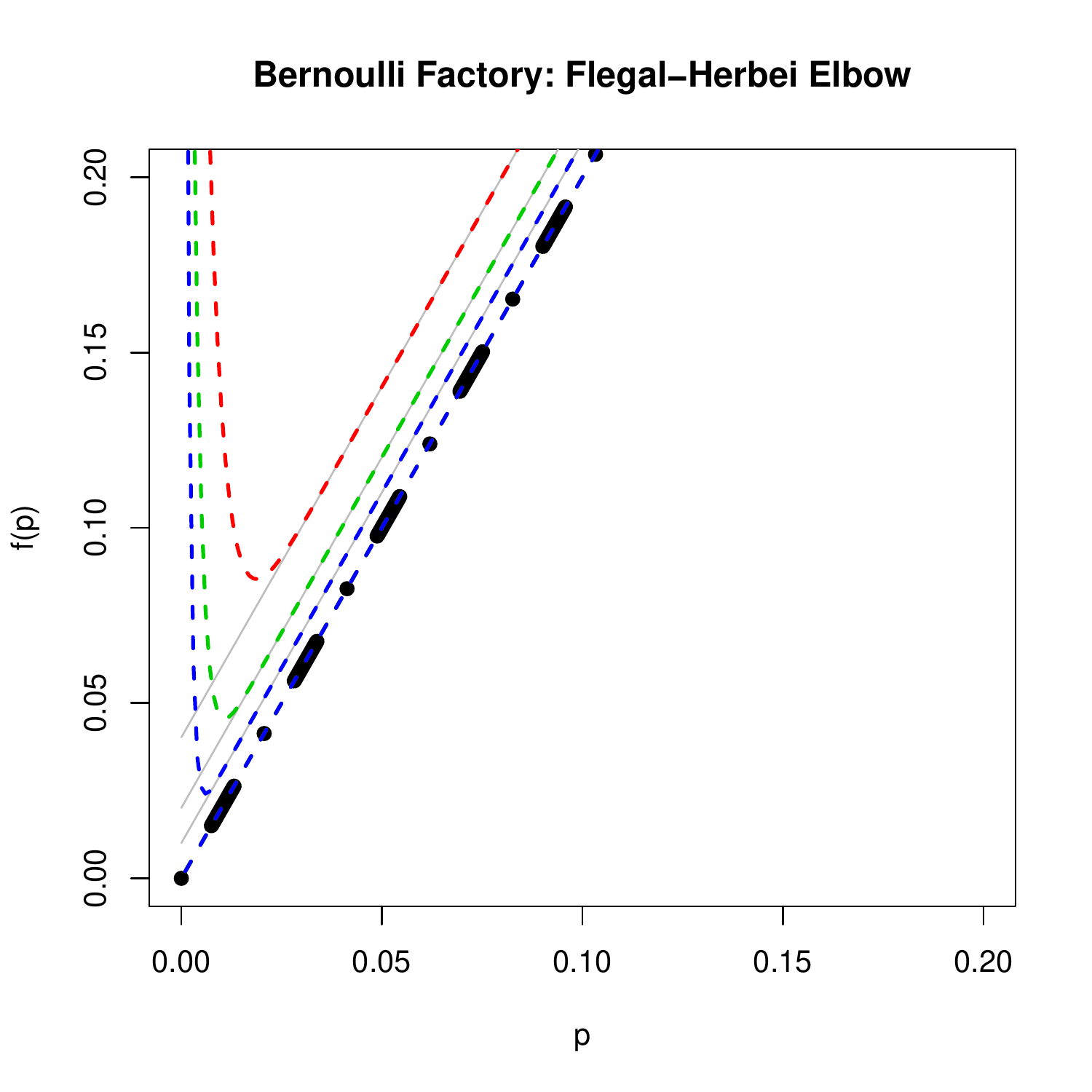}
\includegraphics[width=0.5\linewidth]{\imloc 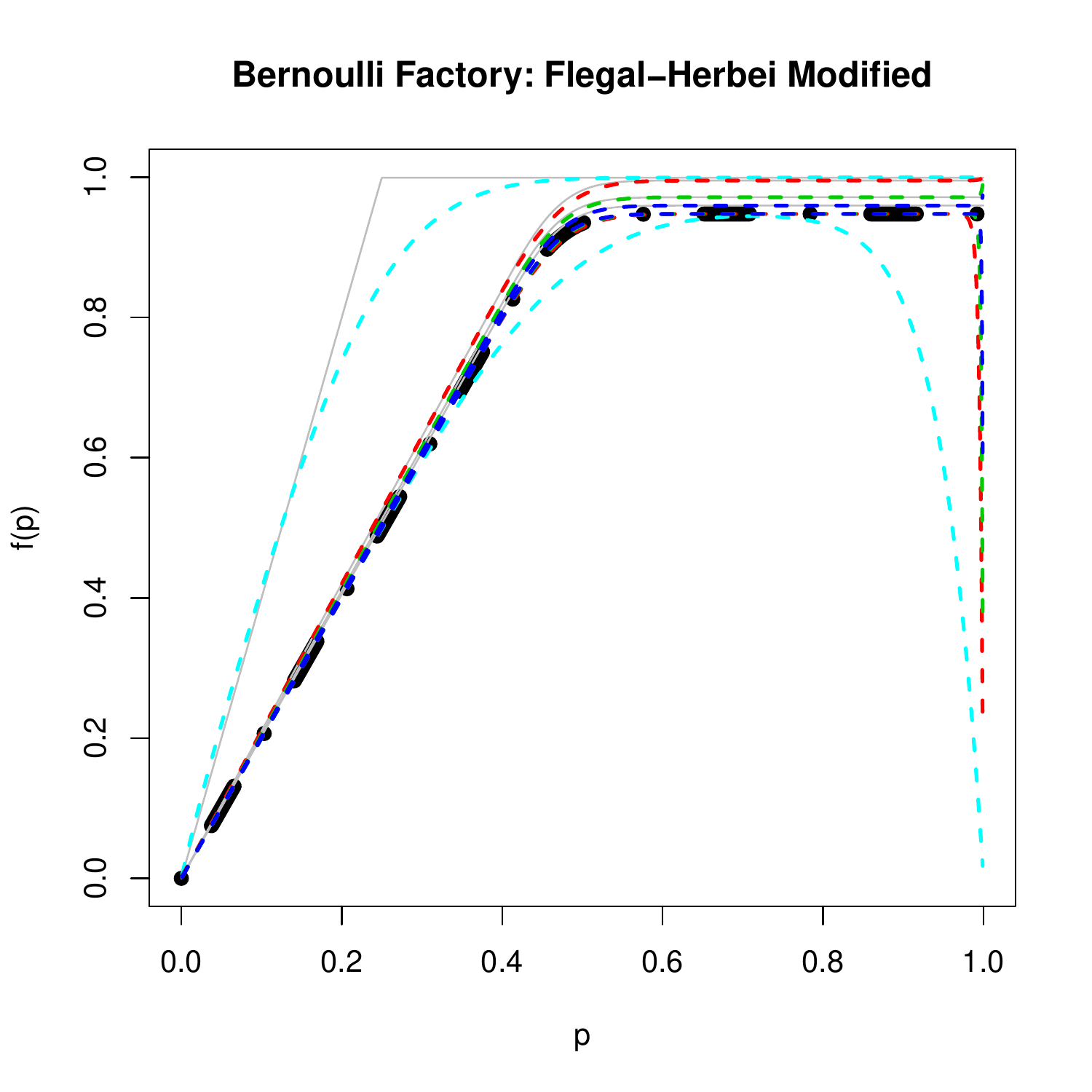}\includegraphics[width=0.5\linewidth]{\imloc 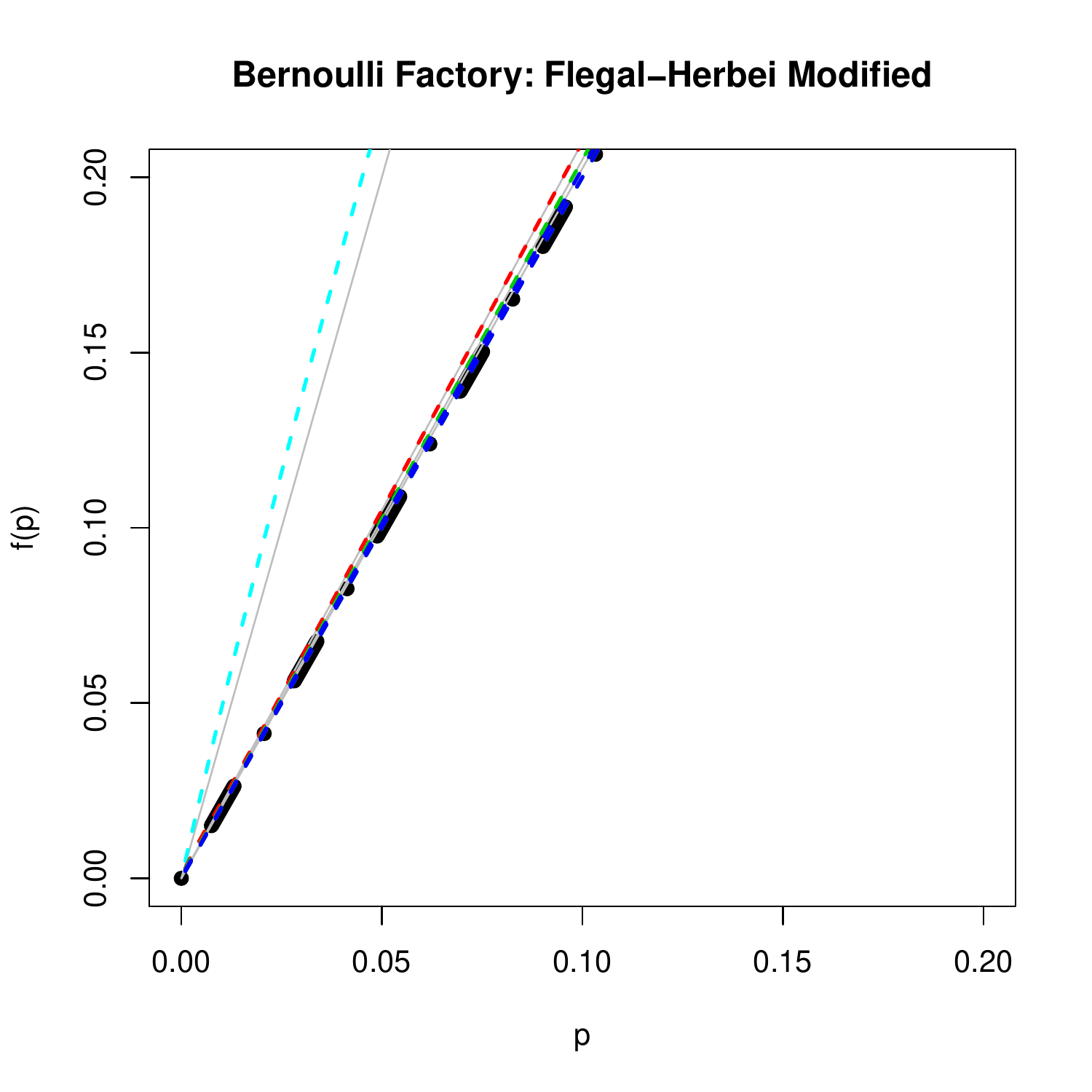}
\caption{The Flegal-Herbei smoothed elbow function, with the original implementation (top) and an envelope-based modification (bottom). The improved version considerably speeds up the algorithm, particularly for small $p$, as the function is additionally optimized for that region of the curve. Additionally, while the first checkpoint of the original algorithm has 256 bits (512 and 1024 follow), an additional envelope pair is added with only 20 bits that allows for even greater speedup.\label{fh-revisited}}
\end{center}
\end{figure}

Figure \ref{fh-revisited} illustrates these modifications with respect to $(c,\epsilon, \delta)=(2,0.2,1/6)$. These envelopes lie closer to the target function for the same number of input bits. The ``preface'' envelope, taking 20 input bits, also substantially cuts down on the running time without compromising the integrity of the original algorithm.

\subsection{Other Differentiable Functions}\label{s:smooth-ones}

Consider a parabolic factory function

\[ f(p) = c\left(1-4\left(p-0.5\right)^2\right), 0 < c < 1 \]
and the square root function

\[ f(p) = \sqrt{p} \]
which are concave and bounded on the interval $(0,1)$. Because of this concavity, the only supplemental envelopes that are required are above the function.

\begin{figure}
\begin{center}
\includegraphics[width=0.5\linewidth]{\imloc 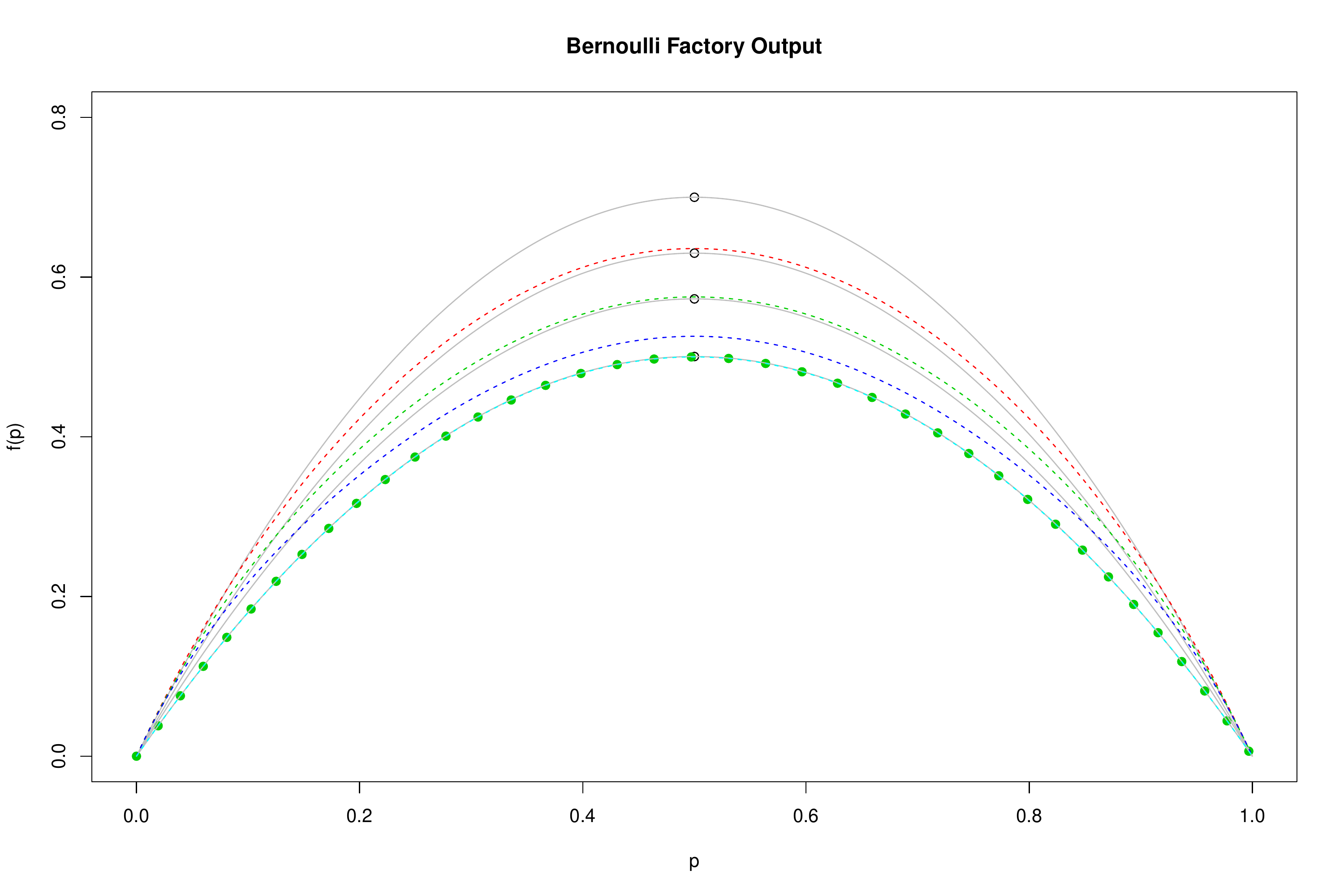}
\caption{An envelope cascade for the factory function $f(p) = c(1-4(p-0.5)^2)$ for $c=0.5$; the target function is the thick green dotted line, while thinner dashed lines are the envelopes (top only shown). The cascade is formed starting with $c_1=0.75$, with each successive $c_i$ set where the previous Bernstein expansion intersected the line $x=0.5$. The convexity argument for the piecewise linear case does not apply here, meaning that the envelope condition must be checked manually.\label{parab-cascade}}

\includegraphics[width=0.5\linewidth]{\imloc 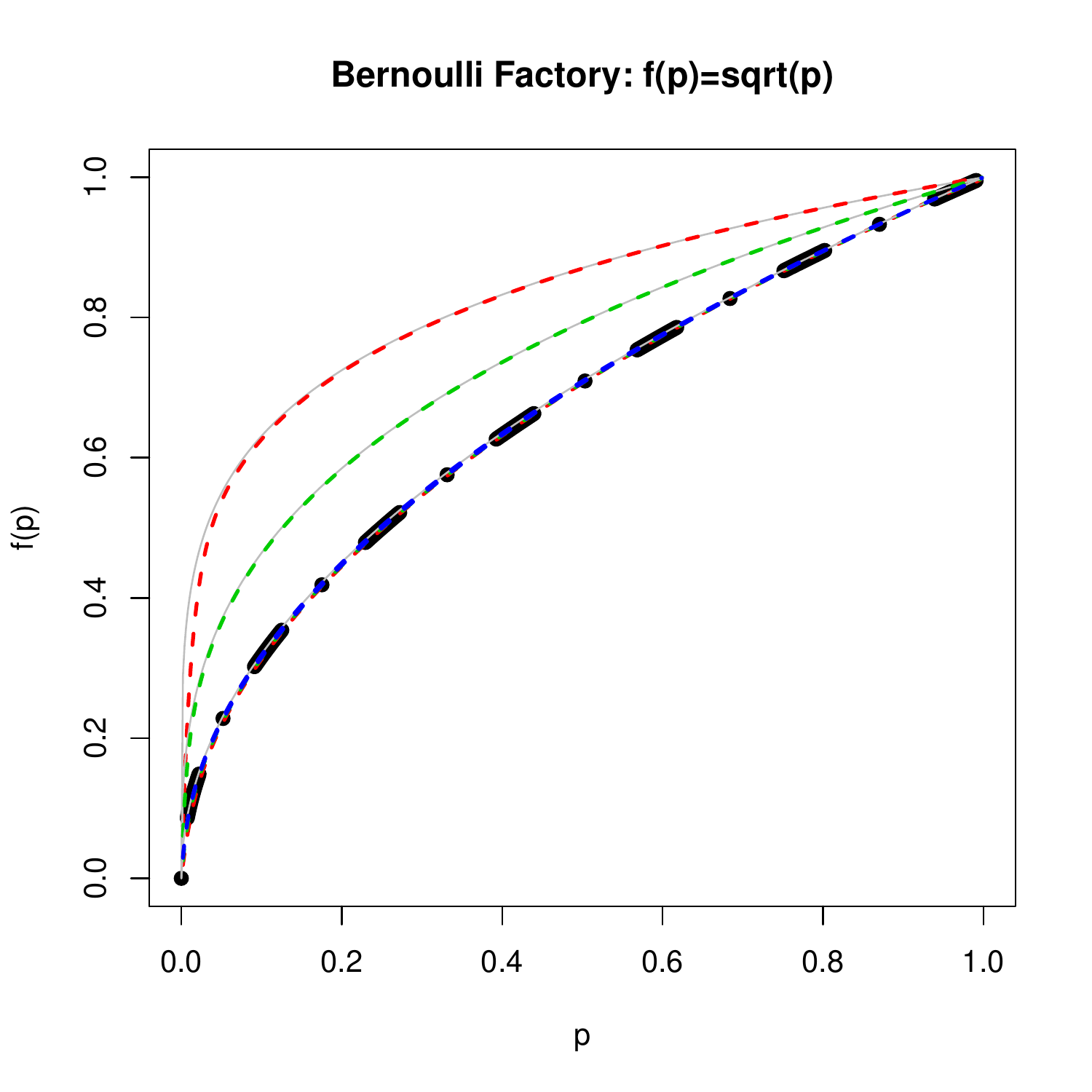}\includegraphics[width=0.5\linewidth]{\imloc 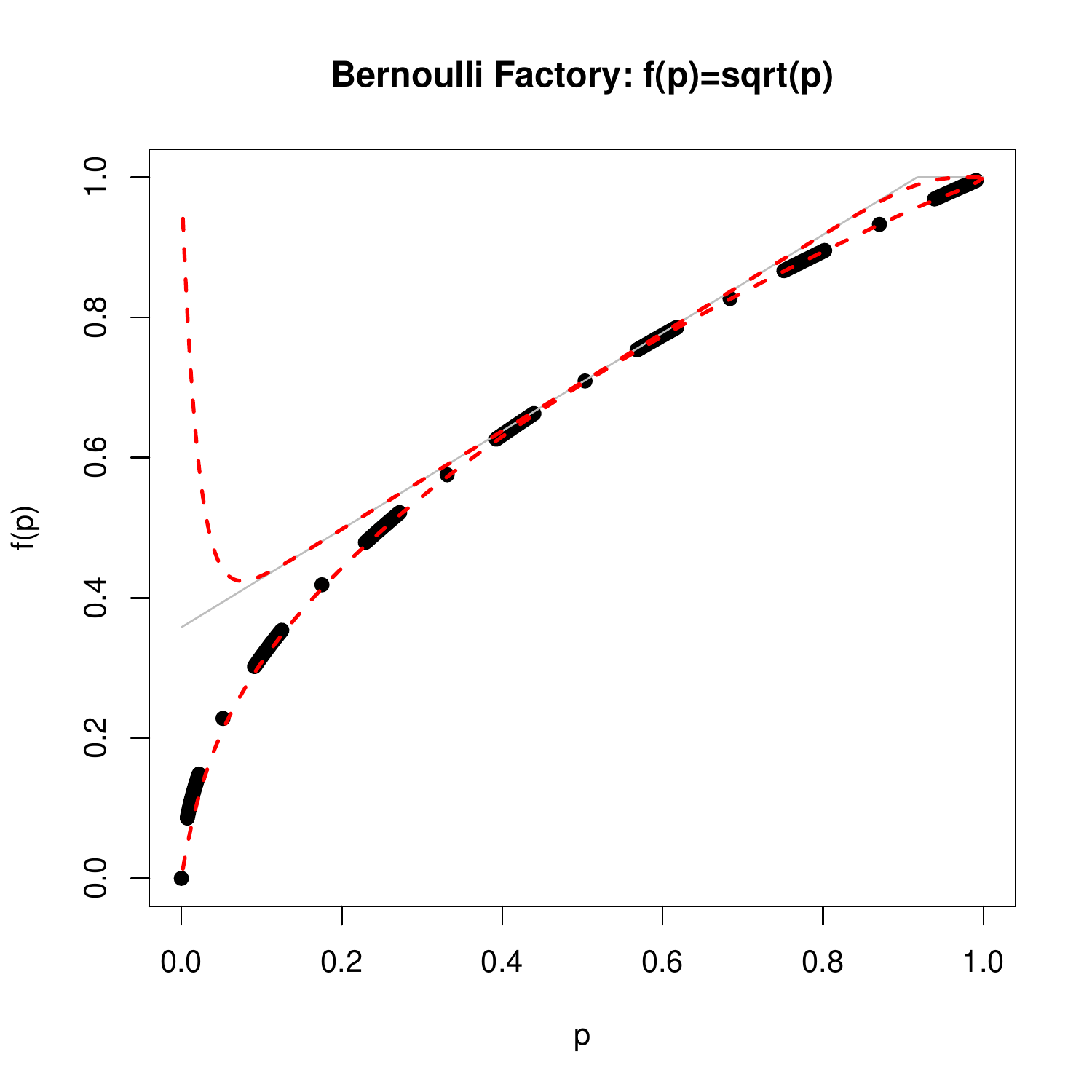}
\caption{Two choices of envelopes for the factory function $f(p) = \sqrt{p}$. Left, a cascade with $h^m(p) = p^{1/q_m}$, where $n=(100,200,300)$ and $q=(5, 3, 2.02)$, for which the envelope conditions are shown to hold manually. Right, a single envelope designed to lie nearly tangent to the target from above, $h^1(p)=\min(0.358+0.7p, 1)$ for $n=50$.\label{sqrt-cascades}}

\end{center}
\end{figure}


Figure \ref{parab-cascade} demonstrates an envelope scheme for the parabolic target in which the envelopes are formed by taking values of $c_i$ that approach the target $c$ from above, by setting the next $c_i$ to be where the previous Bernstein expansion crosses the line $x=0.5$. Manual checking of each set size, with a finite number of elements, guarantees that the envelope condition is satisfied. Because we do not have a piecewise linear target function, we cannot easily determine whether or not the upper envelope's Bernstein approximation is greater than or equal to the target function across the entire interval $[0,1]$ and must check the condition manually. 

Figure \ref{sqrt-cascades} shows how a choice of envelope functions may be made with differing objectives. On the left is a sequence of envelopes designed to approximate the entire target function, by taking power functions that approach $f(p)=p^{0.5}$ from above; on the right, a single envelope function is shown that lies nearly tangent to the factory function at $p=0.5$. If we have knowledge that the input bits have probability in that neighborhood, then this function will be far more efficient than the standard envelope sequence alone. To ensure a balance of efficiency, we can also design a cascade that starts with the tangent envelope, followed by a set of power envelopes that lie between the tangent and the factory function target.

The convergence time properties of these methods are unknown; however, they can be shown to terminate in finite time, since for any $c_i$ there must exist a finite $n$ under which the upper envelope is greater than the target (as the Bernstein polynomial to the upper envelope converges to the envelope from below), so that we can always choose a cascade series $\{c_i\}$ that will converge toward the target function from above. Because these functions both have finite second derivatives on the open interval $(0,1)$, there exists an envelope cascade across the whole domain where the expected time to convergence for any particular $p$ decays as $1/n$, though in the case of the square root, this convergence will be much slower near zero where the second derivative approaches infinity.



\subsubsection*{Acknowledgements}

We thank Peter Glynn, Jim Hobert, Xiao-Li Meng and Christian Robert for their introduction of the problem to us, and James Flegal, Krzysztof Latuszynski and Serban Nacu for various discussions.

\subsubsection*{Supplemental Code}

\begin{description}

\item[R code:] rberfac-public.R contains the information to replicate 
the Bernoulli factory algorithm. (rberfac-public.zip)

\end{description}


\bibliographystyle{\baseloc ims}
\bibliography{\baseloc actbib}


\end{document}